# Broadband three-mode converter and multiplexer based on cascaded symmetric Y-junctions and subwavelength engineered MMI and phase shifters

David González-Andrade[a,*], Irene Olivares[b], Raquel Fernández de Cabo[c], Jaime Vilas[b], Antonio Dias[b], Aitor V. Velasco[c]

[a] *Centre de Nanosciences et de Nanotechnologies, CNRS, Université Paris-Saclay, Palaiseau 91120, France*
[b] *Alcyon Photonics S.L., Madrid 28004, Spain*
[c] *Instituto de Óptica Daza de Valdés, Consejo Superior de Investigaciones Científicas (CSIC), Madrid 28006, Spain*
\* Corresponding author: david.gonzalez-andrade@c2n.upsaclay.fr

ABSTRACT

Mode-division multiplexing has emerged as a promising route for increasing transmission capacity while maintaining the same level of on-chip integration. Despite the large number of on-chip mode converters and multiplexers reported for the silicon-on-insulator platform, scaling the number of multiplexed modes is still a critical challenge. In this paper, we present a novel three-mode architecture based on multimode interference couplers, passive phase shifters and cascaded symmetric Y-junctions. This architecture can readily operate up to the third-order mode by including a single switchable phase shifter. Moreover, we exploit subwavelength grating metamaterials to overcome bandwidth limitations of multimode interference couplers and phase shifters, resulting in a simulated bandwidth of 161 nm with insertion loss and crosstalk below 1.18 dB and -20 dB, respectively.

## 1. Introduction

The relentless growth of global Internet traffic has been driven in recent years by the emergence of data-hungry services and their mass adoption by an increasingly interconnected society [1-3]. Moreover, the cloud nature of many new applications such as machine learning or artificial intelligence require large data sets to be processed on internal servers or transferred between data centers. This resource-intensive paradigm for accessing, computing, and storing data has led to the creation of hyperscale data centers consisting of thousands of servers located in the same physical facility [4]. To cope with the resulting zetta scale of annual data flow, modern data centers have been relying on optical technologies for both long-haul and few-meter interconnects. Compared to their electronic counterparts, these optical technologies offer higher processing speeds, broader bandwidths and lower latency and energy consumption. Silicon photonics, leveraging the mature fabrication facilities of the microelectronics industry, plays a key role in the optical interconnect industry due to its capacity for high-yield and low-cost mass production of high-performance optoelectronic circuits [5,6].

However, the development of next-generation datacenters for Tbps communications and exascale computing systems is not feasible by scaling infrastructures alone and requires increasingly efficient optical interconnects for short-reach distances [7]. As single-mode transmission approaches its fundamental limits, space-division multiplexing has emerged as a promising way to further improve the transmission capacity of optical interconnects through the use of multicore or multimode waveguides [8]. The latter, which is also called mode-division multiplexing (MDM), has attracted an increasing interest as it leverages the orthogonality of the eigenmodes supported by a single multimode waveguide, thus allowing to maintain the same level of on-chip integration [9,10]. That is, MDM enables encoding different data channels into specific spatial modes, increasing capacity proportionally to the number of modes used.

Numerous on-chip mode converters and multiplexers/demultiplexers (MCMD) have been proposed for the silicon-on-insulator (SOI) platform to date. Asymmetric Y-junctions are based on the principle of mode evolution in adiabatic structures, which results in broad operating bandwidths but also in long device lengths [11-13]. The minimum feature size of current lithography processes also has a significant impact in these devices, since the finite resolution at which the tip can be fabricated severely hampers their performance. Asymmetric directional couplers (ADCs) [14], relying on evanescent coupling between adjacent waveguides, are well suited for implementing high-channel count MDM systems, but they typically exhibit narrow bandwidths, and their performance is highly susceptible to fabrication errors. Adiabatic tapers have been employed in the coupling region of ADCs to improve the bandwidth and the resilience against fabrication deviations [15]. MCMDs building upon multimode interference (MMI) couplers and other auxiliary components such as phase shifters (PSs) and symmetric Y-junctions have been proposed as well [16,17], yielding low losses and low crosstalk over a relatively broad wavelength range (~100 nm).

The patterning of silicon at the subwavelength scale has proven to be a simple yet powerful tool to tailor the medium optical properties while inhibiting diffractive effects [35]. More specifically, subwavelength (SWG) metamaterials can behave as a homogeneous metamaterial that provides flexible dispersion and anisotropy engineering, non-feasible in conventional strip and rib waveguides. These properties have led to the realization of Si devices with unprecedented performance over the past 15 years [36-38]. In the MDM field, MCMDs based on subwavelength pixelated structures have demonstrated ultra-compact footprints [18]. SWGs have also been applied to ADCs and triple-waveguide couplers to improve fabrication tolerances and extend the operation bandwidth of conventional counterparts [19-21]. Furthermore, low losses and low crosstalk within ultra-broad bandwidths have also been reported using subwavelength engineered MMI couplers and PSs, and SWG-slot-assisted adiabatic couplers [22-26].

Despite the large number of available two-mode MCMDs, scaling the number of multiplexed modes beyond the fundamental and first-order modes is of great importance to multiply the capacity of next-generation datacom systems. Although it is fairly straightforward to extend operation to a larger number of modes in asymmetric Y-junctions and conventional and tapered ADCs [27], three- and four-mode MCMD based on MMI couplers have only recently been reported [28-32]. However, the proposed architectures are still limited by narrow operating bandwidths and high crosstalk values.

In this work, we propose a novel MCMD architecture based on a 4×4 MMI, three phase shifters and four symmetric 1×2 Y-junctions arranged in a conventional cascaded configuration. The device operates as a three-mode MCDM with passive phase shifters but can readily convert up to the third-order mode by including a single switchable phase shifter. Moreover, we demonstrate loss and crosstalk reduction in a broad bandwidth by SWG-engineering of both the MMI coupler and phase shifters. Simulations show operation bandwidth of 161 nm with insertion loss and crosstalk below 1.18 dB and -20 dB, respectively.

## 2. Principle of operation and device design

To explain the operation principle and the device design, let us first focus on the nanophotonic structure shown in Fig. 1(a) consisting of a conventional 4×4 MMI, three phase shifters (PS1, PS2 and PS3) and four symmetric 1×2 Y-junctions (three identical Y1 and a different one Y2). SWG enhancement of the proposed architecture, shown in Fig. 1(b) and Fig. 1(d) will be discussed in epigraphs 4 and 5. An SOI platform with a thin Si wire surrounded by $SiO_2$ bottom layer and upper cladding are considered. A schematic view of the waveguide cross-section is shown in Fig. 1(c) for clarity.

In order to illustrate the operation of the MCMD, let us focus on the mode evolution and phase relations in each individual constituent of the MCMD. Here, we aim at mode conversion and multiplexing of the first four modes for transverse-electric (TE) polarization, that is, the fundamental mode (TE$_0$), the first-order mode (TE$_1$), the second-order mode (TE$_2$) and the third-order mode (TE$_3$).

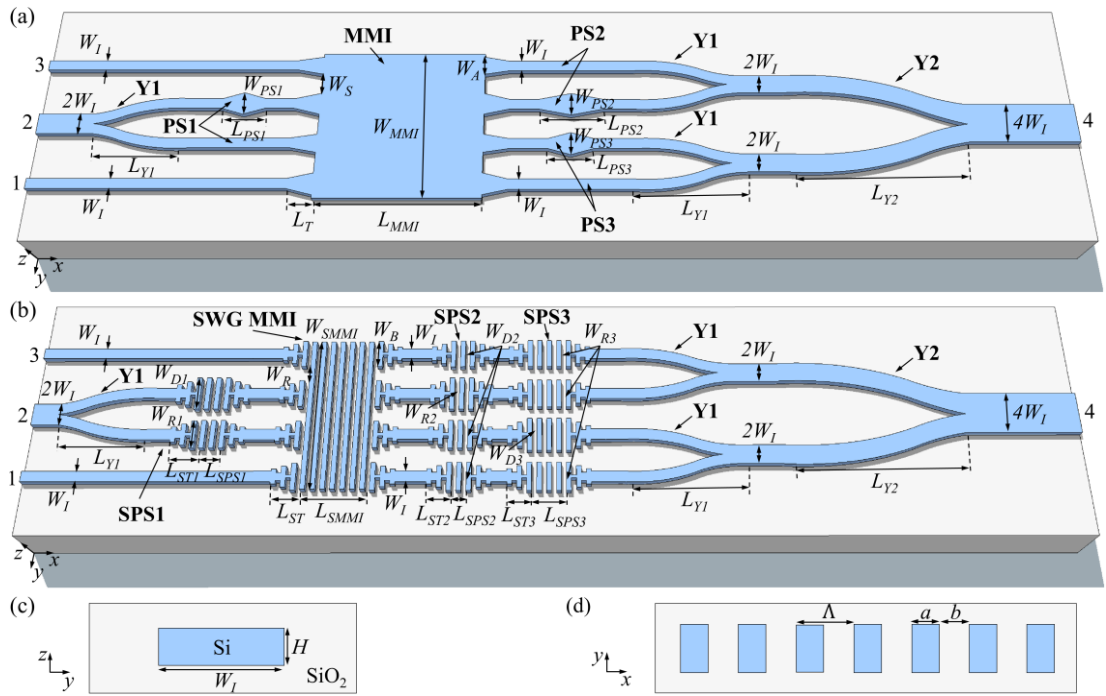

**Fig. 1.** Three-dimensional schematic of the proposed three-mode converter and multiplexer/demultiplexer comprising a 4×4 MMI, three phase shifters and four symmetric Y-junctions implemented with (a) conventional homogeneous and (b) SWG metamaterial waveguides. (c) Cross-sectional view of the SOI strip waveguides with a $SiO_2$ cladding. (c) Top view of the SWG waveguides with their main geometrical parameters.

Our MCMD includes two types of symmetric multimode 1×2 Y-junctions: Y1, with a stem supporting up to two modes; and Y2, with a wider stem supporting up to four modes. In general, multimode symmetric 1×2 Y-junctions transform the two in-phase $m^{th}$-order modes in the arms into the $(2m)^{th}$-order mode in the stem when $m$ is even, and into the $(2m+1)^{th}$-order mode in the steam when $m$ is odd [33]. Likewise, two anti-phase $m^{th}$-order modes in the arms are transformed into the $(2m+1)^{th}$-order mode in the stem when $m$ is even, and into the $(2m)^{th}$-order mode in the stem when $m$ is odd. Figure 2(a) illustrates how this principle affects Y1 operation. Since only two modes are supported by the Y1 stem, a TE$_0$ (red) mode at the stem results in two in-phase TE$_0$ modes at the arms, whereas TE$_1$ (orange) mode at the stem results in two anti-phase TE$_0$ modes at the arms. Figure 2(b) shows the extension of this behavior to four mode operation in Y2. Operation for TE$_0$ (red) and TE$_1$ (orange) is the same as in Y1, whereas injection of TE$_2$ (green) and TE$_3$ (purple) modes through the stem waveguide generates two anti-phase TE$_1$ or two in-phase TE$_1$ modes at the arms, respectively. Therefore, by cascading Y1 and Y2, and judiciously tailoring the phase relations induced by the rest of the MCMD, mode conversion and multiplexing between up to four modes can be achieved. We will hence study the phase shift induced by the 4×4 MMI coupler, and subsequently design a phase shifter architecture that satisfies the phase distributions imposed by the cascaded Y-junctions.

Bachmann et al. already derived a set of equations to calculate the phase relations of $N \times N$ MMI couplers [34]. At this point, it is important to mention that the definition of the phase in this work is $\varphi = \beta x - \omega t$, where $\beta$ is the phase constant (also known as propagation constant), $x$ is the propagation direction and the term $-\omega t$ corresponds to the temporal dependence. As in [34] the authors used the opposite phase convention, i.e., $= \omega t - \beta x$, equations can be rewritten as follows:

$$i + j \text{ even: } \varphi_{ij} = -\varphi_0 - \pi - \frac{\pi(j-i)(2N-j+i)}{4N}, \quad (1)$$

$$i + j \text{ odd: } \varphi_{ij} = -\varphi_0 - \frac{\pi(j+i-1)(2N-j-i+1)}{4N}, \quad (2)$$

where $\varphi_0$ is a constant phase, $i$ and $j$ are the indices of the $N$ inputs and outputs, respectively. Using Eqs. (1) and (2), the phase relations of a 4×4 MMI coupler can be calculated as shown in Table 1. Please note the input and output numbering in Fig. 3.

**Table 1**
Calculated phase relations $\boldsymbol{\varphi_{ij}}$ of a 4×4 MMI coupler.

| $i$ \ $j$ | 1 | 2 | 3 | 4 |
|---|---|---|---|---|
| 1 | $-\pi$ | $-3\pi/4$ | $-7\pi/4$ | $-\pi$ |
| 2 | $-3\pi/4$ | $-\pi$ | $-\pi$ | $-7\pi/4$ |
| 3 | $\pi/4$ | $-\pi$ | $-\pi$ | $-3\pi/4$ |
| 4 | $-\pi$ | $\pi/4$ | $-3\pi/4$ | $-\pi$ |

We then calculate, for each input port, the resulting phase difference at the two upper output ports ($\Delta\varphi_{12}$) and the two lower ports ($\Delta\varphi_{34}$) as:

$$\Delta\varphi_{12} = \varphi_{i1} - \varphi_{i2}, \quad (3)$$
$$\Delta\varphi_{34} = \varphi_{i3} - \varphi_{i4}, \quad (4)$$

Calculated phase differences are shown in Table 2. Since phase evolution at both the MMIs and Y-splitters are fixed, we then need to design a combination of PSs (placement and phase shift values), that results in the required phase relations. As shown in Figure 1(a), we achieve this goal by including a first phase shifter (PS1) between inputs 3 and 2 of the MMI, with a phase shift of $\pi/2$; a second phase shifter (PS2) between outputs 2 and 1, with a phase shift of $-\pi/4$; and a third phase shifter (PS3) between outputs 3 and 4 with a phase shift of $3\pi/4$. An additional two-mode Y-junction (Y1) is included at MCMD port 2 to satisfy even-order modes phase conditions, as discussed hereunder.

Figure 4 shows the operation of the device working in multiplexer configuration, including the value of the phase relations at different locations for clarity. It should be noted that phase values have been calculated with with respect to the mode phase at the input ports, which is considered to be zero for simplicity. When light is injected through MCMD port 1 [Fig. 4(a)], the combination of all the aforementioned phase relations results in all modes arriving in-phase at the arms of the cascaded Y-junctions. Thus, two in-phase $TE_0$ modes are coupled into Y1 stems, which subsequently generate the desired $TE_0$ mode at the multimode stem waveguide of Y2 (MCMD port 4).

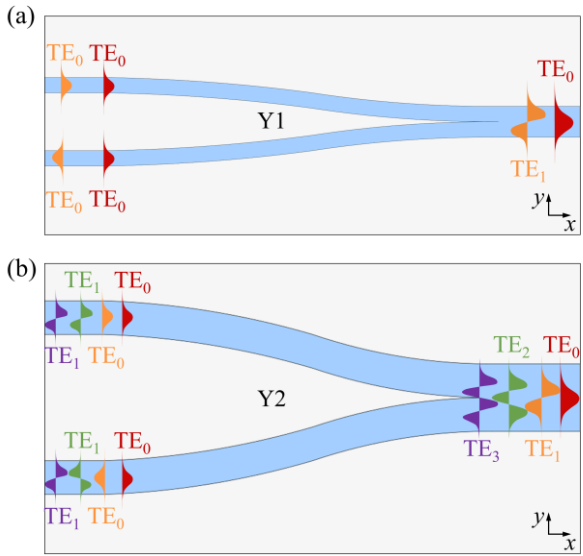

**Fig. 2.** Schematic and principle of operation of a multimode symmetric 1×2 Y-junction for (a) a two-mode stem, and (b) a four-mode stem.

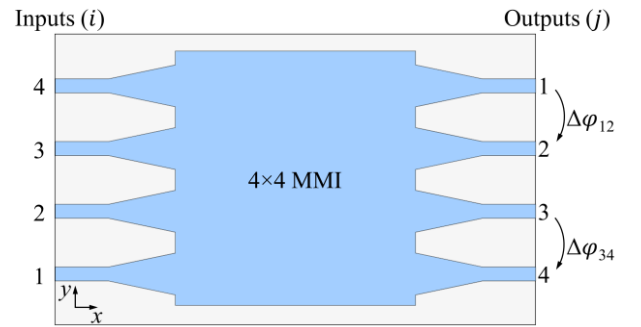

**Fig. 3.** Schematic of a 4×4 MMI coupler, illustrating port and phase notations.

**Table 2**
Calculated phase differences between MMI output pairs for each input.

| $i$ | $\Delta\varphi_{12}$ | $\Delta\varphi_{34}$ |
|---|---|---|
| 1 | $-\pi/4$ | $-3\pi/4$ |
| 2 | $\pi/4$ | $3\pi/4$ |
| 3 | $5\pi/4$ | $-\pi/4$ |
| 4 | $-5\pi/4$ | $\pi/4$ |

When light is injected through MCMD port 2 [Fig. 4(b)], the combination of Y1 and PS1 results in simultaneous light coupling to MMI input ports 2 and 3, but with a $\pi/2$ phase difference. This in turn generates in-phase modes in the upper arms that are in anti-phase with the two in-phase modes in the lower arms at their arrival at the cascaded Y-junctions. This combination results in $TE_1$ generation at the MCMD output.

Finally, when light is injected through MCMD port 3 [Fig. 4(c)], that is, MMI input port 4, in-phase modes are generated in the middle arms, which are in anti-phase with the two in-phase modes generated in the top and bottom arms, before the cascaded Y-junctions. This results in anti-phase $TE_1$ modes at the output of Y1 stems, which subsequently generate the $TE_2$ mode at the MCDM output.

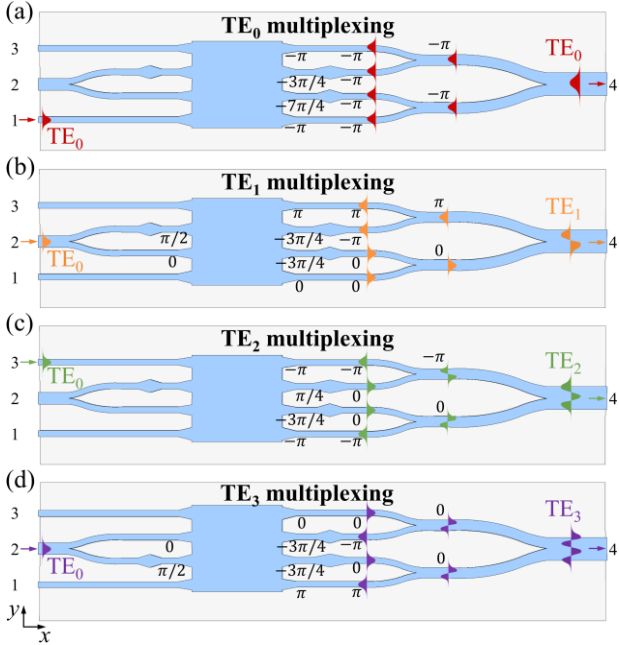

**Fig. 4.** Principle of operation of the proposed three-mode converter and multiplexer/demultiplexer for (a) $TE_0$, (b) $TE_1$, (c) $TE_2$ and (d) $TE_3$ mode multiplexing.

So far, we have only considered passive PSs, that is, PSs with a fixed phase shift. However, if the phase introduced by PS1 is $3\pi/2$ instead of $\pi/2$, it is possible to generate the $TE_3$ mode at MCMD output [Fig. 4(d)]. For illustrative purposes, we represent this phase shift by switching the position of the tapers in PS1. This feature opens the possibility of extending MCMD operation to four modes using a single switchable PS.

**3. Proof-of-concept results**

To verify the principle of operation explained in the previous section, we firstly optimized each constituent (i.e., MMI, phase shifters and Y-junctions) for a design wavelength of 1550 nm. We chose a standard silicon thickness of $H = 220$ nm and an interconnection waveguide width of $W_I = 400$ nm. Thus, symmetric Y-junctions are designed with stem widths of $2W_I = 800$ nm for Y1 and $4W_I = 1600$ nm for Y2. Geometrical parameters of the 4×4 MMI coupler, the phase shifters and the symmetric Y-junction are summarized in Table 3.

In order to evaluate the performance of each constituent element, the figures of merit for the MMI are the excess loss (EL), imbalance (IB) and phase error (PE):

$$EL_i\ [dB] = -10\log_{10}\left(\sum_j |S_{ji}|^2\right), \quad (5)$$

$$IB_i^{jk}\ [dB] = 10\log_{10}\left(|S_{ji}|^2/|S_{ki}|^2\right), \quad (6)$$

$$PE_i^{jk}\ [°] = [\angle(S_{ji}/S_{ki}) - \varphi_{ideal}] \cdot 180/\pi, \quad (7)$$

where $S_{ji}$ and $S_{ki}$ are the scattering parameters for input $i$ and outputs $j$ and $k$, and $\varphi_{ideal}$ is the ideal phase relation depending on selected input and output ports as shown in Table 1. The designed 4×4 MMI exhibits EL <

**Table 3**
Geometrical parameters of the three-mode converter and multiplexer/demultiplexer with homogeneous waveguides.

| Constituent | Parameter | | Value |
|---|---|---|---|
| Waveguides | Width | $W_I$ | 400 nm |
| MMI | Separation | $W_S$ | 500 nm |
| | Access width | $W_A$ | 1.3 µm |
| | Taper length | $L_T$ | 6 µm |
| | MMI width | $W_{MMI}$ | 7.2 µm |
| | MMI length | $L_{MMI}$ | 91 µm |
| Y1 | Arm width | $W_I$ | 400 nm |
| | Arm length | $L_{Y1}$ | 5 µm |
| | Stem width | $2W_I$ | 800 nm |
| Y2 | Arm width | $2W_I$ | 800 nm |
| | Arm length | $L_{Y2}$ | 20 µm |
| | Stem width | $4W_I$ | 1.6 µm |
| PS1 | PS width | $W_{PS1}$ | 600 nm |
| | PS length | $L_{PS1}$ | 2.41 µm |
| PS2 | PS width | $W_{PS2}$ | 600 nm |
| | PS length | $L_{PS2}$ | 8.38 µm |
| PS3 | PS width | $W_{PS3}$ | 600 nm |
| | PS length | $L_{PS3}$ | 3.61 µm |

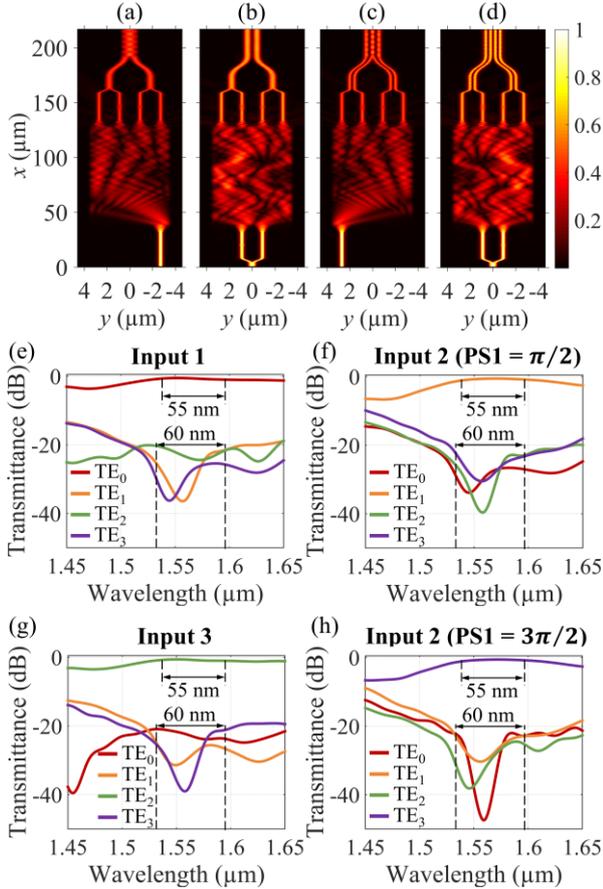

**Fig. 5.** Electric field amplitude $|E|$ in the XY plane at the middle of the silicon layer for (a) $TE_0$, (b) $TE_1$, (c) $TE_2$ and (d) $TE_3$ mode multiplexing. Simulated transmittance to output port 4 as a function of the wavelength when $TE_0$ mode is launched into (e) input port 1, (f) input port 2 with $PS1 = \pi/2$, (g) input port 3 and (h) input port 2 with $PS1 = 3\pi/2$. Vertical lines indicate the bandwidth where IL < 1 dB (55 nm) and XT < −20 dB (60 nm) are achieved for all modes simultaneously.

0.54 dB, IB < ±0.4 dB and PE < ±0.32° at the wavelength of $\lambda_0 = 1550$ nm. Regarding the spectral response, EL < 2.15 dB, IB < ±8.1 dB and PE < ±46.03° are attained in the entire simulated wavelength range (1.45 – 1.65 µm).

Designed phase shifters introduce small phase deviations of only 0.12° for PS1, 0.13° for PS2 and 0.16° for PS3, with respect to their target phase difference at 1550 nm. However, considering the simulated bandwidth of 200 nm, phase errors increase up to 9.84° for PS1, 22.28° for PS2, and 12.12° for PS3.

Symmetric Y-junctions Y1 and Y2 were also designed showing negligible excess losses and power imbalance between output ports at the design wavelength. More specifically, Y1 losses are lower than 0.01 dB for both $TE_0$ and $TE_1$ mode operation in the 1.45 – 1.65 µm wavelength range. Conversely, calculated excess losses for Y2 are below 0.15 dB for $TE_0$, $TE_1$, $TE_2$ and $TE_3$ mode operation within the same bandwidth.

Once all elements were optimized, two-dimensional finite-difference time-domain (FDTD) simulations of the whole MCMD were performed by applying the effective index method to the original three-dimensional structure [see Fig. 1(a)]. The simulated field distribution of the three-mode MCMD is shown in Fig. 5(a)-(d), demonstrating the successful implementation of the phase relations described in section 2.

Some ripples can be observed for $TE_0$ and $TE_2$ mode multiplexing in the stem waveguide of Y-junction Y2 [see Figs. 5(a) and 5(c)], which we attribute to a higher crosstalk between both modes compared to $TE_1$ and $TE_3$ mode multiplexing.

The transmittance as a function of the wavelength was computed for the complete MCMD [see Fig.5(e)-(h)]. At the central wavelength of $\lambda_0 = 1550$ nm, insertion losses are lower than 0.53 dB, 0.79 dB and 0.59 dB for the generation of $TE_0$, $TE_1$ and $TE_2$ modes in the stem waveguide, respectively. Our device also exhibits a low crosstalk at the same wavelength with values below -21.61 dB for $TE_0$, -28.94 dB for $TE_1$ and -21.11 dB for $TE_2$.

By tuning the value of PS1 to $3\pi/2$, $TE_3$ mode (instead of $TE_1$ mode) can be generated. In this case, insertion losses are below 0.75 dB, and the crosstalk is better than -28.79 dB, both at 1550 nm. These results corroborate the higher crosstalk for $TE_2$ mode operation, which leads to a slight ripple in the field distribution at port 4.

Regarding performance across the spectrum, insertion losses lower than 1 dB are attained for a 55 nm bandwidth (1542 – 1597 nm), whereas the crosstalk is below -20 dB for a 60 nm bandwidth (1537 –1597 nm) as shown with vertical lines in Fig. 5. These results prove the correct operation of the proposed architecture, but the overall bandwidth is significantly limited by the narrow spectral response of both the MMI and PSs.

### 4. SWG performance enhancement

To overcome these bandwidth limitations, we propose the MCMD with SWG metamaterials shown in Fig. 1(b). The design of each of the constituents of the SWG MCMD was performed by individual three-dimensional FDTD simulations. The three symmetric Y-junction labeled Y1 maintain the same geometrical dimensions as those used for the conventional multiplexer for the arm and stem widths (see Table 3), but arm length was shortened to $L_{Y1} = 2$ µm. Y-junction Y2 was slightly redesigned to reduce the crosstalk between $TE_0$ and $TE_2$ modes by increasing the length of the arms to $L_{Y2} = 40$ µm.

A procedure similar to those already reported in [39,40] was followed for the optimization of the 4×4 SWG MMI. We restrict the value of the duty cycle (DC = $a/\Lambda$) to 0.5 in order to maximize the minimum feature size for a given period ($\Lambda$) [see Fig.1(d)]. We explored then different periods and found that $\Lambda = 222$ nm significantly flattens the beat length across the spectrum. Compared to the conventional MMI section design, the width $W_{SMMI}$ is increased by 0.8 µm but the length $L_{SMMI}$ is reduced by more than half to ~41.3 µm. To increase the quality of the interferometric patterns formed in the MMI, the access width is $W_B = 1.7$ µm and the separation is reduced to $W_R = 0.3$ µm. The transition between the interconnection waveguides ($W_I = 400$ nm) and the access to the MMI section ($W_B$) is performed by means of adiabatic SWG tapers with a length $L_{ST} = 13.32$ µm. The performance of the 4×4 SWG MMI is shown in Fig. 6(a)-(c). Owing to the symmetry of the structure, only the results obtained when injecting light into input ports 1 and 2 are depicted. It is observed that the device exhibits EL < 0.77 dB, IB < ±1 dB and PE < ±8.02° within a broad bandwidth of 200 nm (1.45 – 1.65 µm).

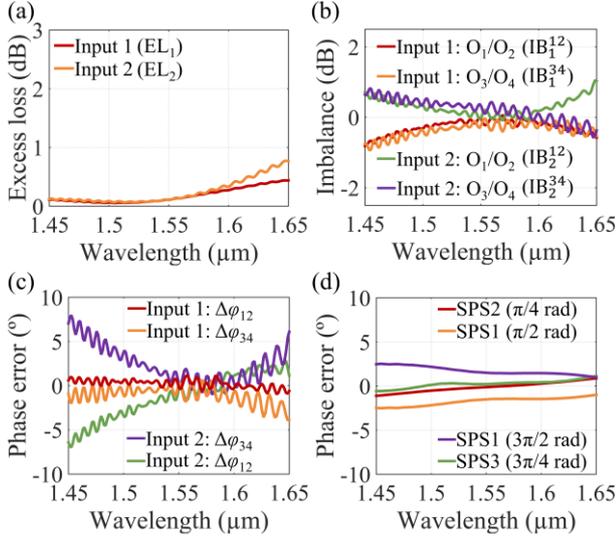

**Fig. 6.** Simulated performance of the 4×4 SWG MMI including (a) excess loss, (b) imbalance and (c) phase error between output ports. (d) Phase error of each SWG PSs as a function of the wavelength.

To drastically extend the operating bandwidth of the nanophotonic phase shifters, we build upon the strategy we recently reported in [41] to develop SWG phase shifters SPS1, SPS2 and SPS3. Notwithstanding, here we employ four parallel SWG waveguides of two different widths to implement SPS2 and SPS3. That is, each PS has three identical reference SWG waveguides with width $W_D$, and one dissimilar SWG waveguide with width $W_R$. Both the reference and dissimilar waveguides have a length of $L_{SPS}$. Note that for SPS1 this configuration is not necessary as only two MMI inputs are illuminated for TE$_1$ and TE$_3$ mode generation. Analogous to the 4×4 SWG MMI, a flat phase shift can be achieved by judicious selecting the SWG period and duty cycle. A duty cycle of 0.5 was fixed to maximize MFS, while a period of 200 nm resulted in minimum phase shift deviation. In order to induce $\pi/4$, $\pi/2$, and $3\pi/4$ phase shifts, we selected respectively $W_{D2} = 1.8$ µm, $W_{R2} = 1.6$ µm, $L_{SPS2} = 6.2$ µm and $L_{ST2} = 3.0$ µm for SPS2; $W_{D1} = 1.8$ µm, $W_{R1} = 1.6$ µm, $L_{SPS1} = 16.8$ µm and $L_{ST1} = 3.0$ µm for SPS1; and $W_{D3} = 1.8$ µm, $W_{R3} = 1.6$ µm, $L_{SPS3} = 28.2$ µm and $L_{ST3} = 3.0$ µm for SPS3. The simulated phase shifts are shown in Fig. 6(d). Negligible deviations can be appreciated with phase shift errors as small as 2.29° for SPS1, and 1.15° for SPS2 and SPS3 within the entire 1.45 – 1.65 µm wavelength range.

## 5. SWG results

The simulation of the entire MCMD is quite resource-intensive and time-consuming due to the device footprint and the need for a fine mesh to simulate SWG-based devices. Thus, instead of performing the full device simulation, we leverage the S-parameter matrices calculated during the design process and concatenate all of them using a circuit simulator to obtain the S-parameter matrix and hence the spectral response of the complete device. The circuit simulator enables bidirectional signals to be accurately simulated, including coupling of modes in the single elements. Figure 7 shows the overall transmittance of the SWG MCMD. Insertion losses (ILs) are lower than 0.37 dB, 0.47 dB and 0.37 dB for $TE_0$, $TE_1$ and $TE_2$ multiplexing, respectively, at the central wavelength of $\lambda_0 = 1550$ nm. Moreover, low crosstalk (XT) is achieved at the same wavelength with values below -21.54 dB for $TE_0$, -32.89 dB for $TE_1$ and -21.24 dB for $TE_2$ multiplexing.

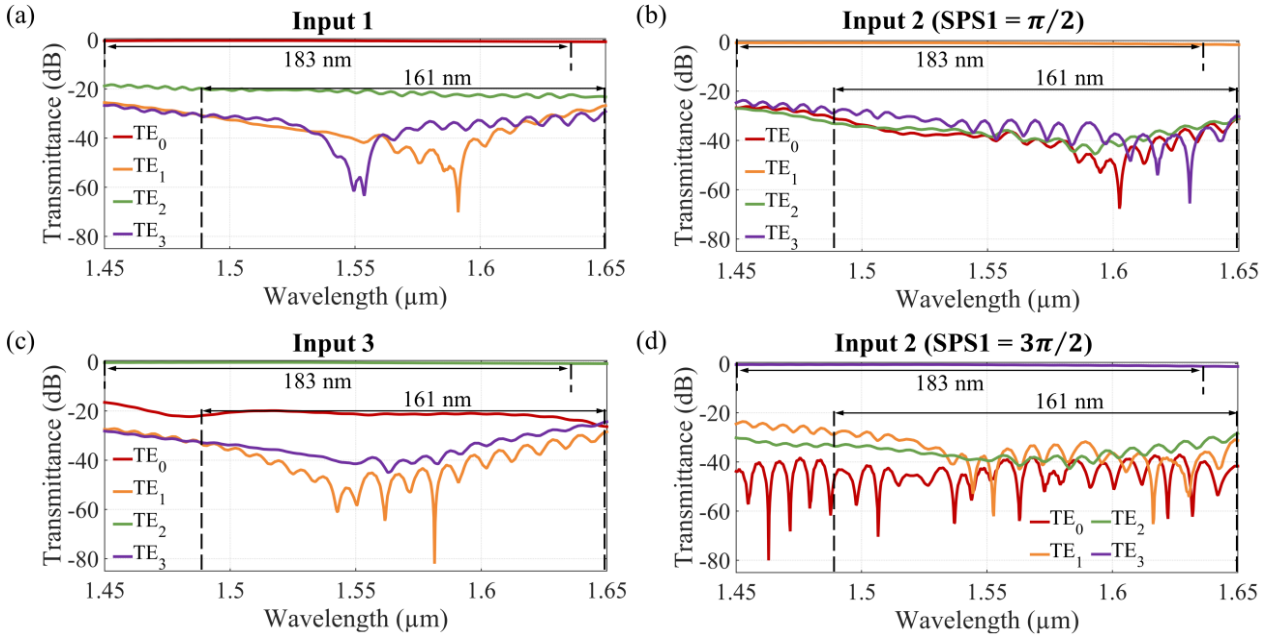

**Fig. 7.** Simulated transmittance as a function of the wavelength of the MCMD with SWG metamaterials when TE$_0$ mode is launched into (a) input port 1, (b) input port 2 with SPS1 = $\pi/2$, (c) input port 3 and (d) input port 2 with SPS1 = $3\pi/2$. Vertical lines indicate the bandwidth where IL < 1 dB (183 nm) and XT < −20 dB (161 nm) are achieved for all modes simultaneously.

**Table 4**
Performance comparison of state-of-the-art three- and four-MCMD based on MMI couplers.

| Ref. | IL [dB] | XT [dB] | BW [nm] | Length [μm] | No. of modes |
|---|---|---|---|---|---|
| [28] | <0.71 | <-18 | 100 | 400 | 3 |
| [29] | <0.9 | <-17 | 40 | 400 | 3 |
| [30] | <1.1 | <-10 | 90 | 17.05 | 3 |
| [31] | <1.3 | <-15 | 40 | 120 | 3 |
| [32] | <3.3 | <-14 | 70 | 173 | 4 |
| [32] | <2.25 | <-15 | 140 | 123 | 4 |
| This work | <1.18 | <-20 | 161 | 181.14 | 3 |

When SPS1 takes the value of $3\pi/4$, insertion losses for $TE_3$ multiplexing reach a low value of 0.47 dB at 1550 nm, while crosstalk values are lower than -39.48 dB for the same wavelength.

This design also shows an excellent performance over a broad bandwidth (BW) of 200 nm with insertion loss lower than 1.18 dB and crosstalk below -16.53 dB. Insertion losses decrease to 1 dB when the bandwidth is restricted to 183 nm (1450 – 1633 nm), whereas a crosstalk below -20 dB is achieved over a 161 nm bandwidth (1489 – 1650 nm). For the sake of comparison, Table 4 summarizes the performance of other three- and four-mode MCMD that are based on MMI couplers and have been reported in the state of the art. To the best of our knowledge, it is the first time such low losses and crosstalk are achieved in an outstanding 161 nm wavelength range.

## 6. Conclusions

In this work, we have proposed a novel architecture to scale the number of multiplexed modes of mode converters and multiplexer based on MMI couplers. Unlike other reported architectures that use unconventional 1×4 Y-junctions or 1×3 Ψ-junctions, here we employ symmetric 1×2 Y-junctions arranged in a conventional cascaded configuration. The design methodology was proposed on the basis of a two-dimensional model with conventional homogenous components (i.e., without patterning the silicon waveguide). The conventional mode converter and multiplexer features sub-decibel insertion loss and crosstalk better than -20 dB in the 1542 – 1597 nm wavelength range. Once the principle of operation was verified, we redesigned and optimized the mode converter and multiplexer by incorporating subwavelength grating metamaterials to leverage the additional degrees of freedom they introduced into the design. A broad design bandwidth of 161 nm for insertion losses below 1.18 dB and crosstalk lower than -20 dB was confirmed by 3D FDTD simulations, comparing very favorably to state-of-the-art three- and four-mode converters and multiplexers. The crosstalk between $TE_0$ and $TE_1$ modes could be further reduced by including optimized Y-junction geometries that mitigate the effect of the non-perfect tip at the junction [42-44]. We believe that our design strategy will open promising prospects for the development of high-performance mode converters and multiplexer based on MMI couplers with a high channel count.

**Credit authorship contribution statement**

**David González-Andrade:** Conceptualization, Methodology, Software, Validation, Formal analysis, Investigation, Data curation, Writing – original draft, Visualization. **Irene Olivares:** Methodology, Software, Validation, Formal analysis, Data curation, Writing – review & editing. **Raquel Fernández de Cabo:** Software, Validation, Data curation, Writing – review & editing. **Jaime Vilas:** Writing – review & editing. **Antonio Dias:** Resources, Writing – review & editing, Project administration, Funding acquisition. **Aitor V. Velasco:** Resources, Writing – review & editing, Supervision, Project administration, Funding acquisition.

**Declaration of Competing Interest**

The authors declare that they have no known competing financial interests or personal relationships that could have appeared to influence the work reported in this paper.

**Acknowledgements**

This work has been funded in part by the Spanish Ministry of Science and Innovation (MICINN) under grants RTI2018-097957-B-C33, PID2020-115353RA-I00; the Spanish State Research Agency (MCIN/AEI/10.13039/501100011033); the Community of Madrid – FEDER funds (S2018/NMT-4326); the European Union – NextGenerationEU through the Recovery, Transformation and Resilience Plan (DIN2020-011488, PTQ2021-011974); and the European Union's Horizon Europe research and innovation program under the Marie Sklodowska-Curie grant agreement Nº 101062518.

**References**

[1] R. Sabella, Silicon photonics for 5G and future networks, IEEE J. Sel. Top. Quantum Electron. 26 (2020) 8301611.

[2] I. Demirkol, D. Camps-Mur, J. Paradells, M. Combalia, W. Popoola, H. Haas, Powering the Internet of things through light communication, IEEE Commun. Mag. 57 (2019) 107-113.

[3] M Filer, J. Gaudette, Y. Yin, Z. Bakhtiari, J. L. Cox, Low-argin optical networking at cloud scale, J. Opt. Commun. Netw. 11 (2019) C94-C105.

[4] Q. Cheng, M. Bahadori, M. Glick, S. Rumley, K. Bergman Recent advances in optical technologies for data centers: a review, Optica 5 (2018) 1354-1370.

[5] D. Thomson, A. Zilkie, J. E. Bowers, T. Komljenovic, G. T. Reed, L. Vivien, D. Marris-Morini, E. Cassan, L. Virot, J.-M. Fédeli, J.-M. Hartmann, J. H. Schmid, D. X. Xu, F. Boeuf, P. O'Brien, G. Z. Mashanovich, M. Nedeljkovic, Roadmap on silicon photonics, J. Opt. 18 (2016) 073003.

[6] X. Chen, M. M. Milosevic, S. Stanković, S. Reynolds, T. Domínguez-Bucio, K. Li, D. J. Thomson, F. Gardes, G. T. Reed, The emergence of silicon photonics as a flexible technology platform, Proc. IEEE 106 (2018) 2101-2116.

[7] S. Bernabé, Q. Wilmart, K. Hasharoni, K. Hassan, Y Thonnart, P. Tissier, Y. Désières, S. Olivier, T. Tekin, B. Szelag, Silicon photonics for terabit/s communication in data centers and exascale computers, Solid State Electron. 179 (2021) 107928.

[8] W. Shi, Y. Tian, A. Gervais, Scaling capacity of fiber-optic transmission systems via silicon photonics, Nanophotonics 9 (2020) 4629-4663.

[9] C. Li, D. Liu, D. Dai, Multimode silicon photonics, Nanophotonics 8 (2018) 227-247.

[10] I. Cristiani, C. Lacava, G. Rademacher, B. J. Puttnam, R. S. Luìs, C. Antonelli, A. Mecozzi, M. Shtaif, D. Cozzolino, L. K. Oxenløwe, J. Wang, Y. Jung, D. J. Richardson, S. Ramachandran, M. Guasoni, K. Krupa, D. Kharenko, A. Tonello, S. Wabnitz, D. B. Philips, D. Faccio, T. G. Eurser, S. Xie, P. St. J. Russell, D. Dai, Y. Yu, P. Petropoulos, F. Gardes, F. Parmiagiani, Roadmap on multimode photonics, J. Opt. 24 (2022) 083001.

[11] J. B. Driscoll, R. R. Grote, B. Souhan, J. I. Dadap, M. Lu, R. M. Osgood, Asymmetric Y junctions in silicon waveguides for on-chip mode-division multiplexing, Opt. Lett. 38 (2013) 1854-1856.

[12] W. Chen, P. Wang, T. Yang, G. Wang, T. Dai, Y. Zhang, L. Zhou, X. Jiang, J. Yang, Silicon three-mode (de)multiplexer based on cascaded asymmetric Y junctions, Opt. Lett. 41 (2016) 2851-2854.

[13] T. A. Tran, H. D. T. Nguyen, C. D. Truong, H. T. Nguyen, Y. V. Vu, D. H. Tran, Three-mode multiplexed device based on tilted- branch bus structure using silicon waveguide, Photonics Nanostructures – Fundam. Appl. 35 (2019) 100709.

[14] D. Dai, J. Wang, Y. Shi, Silicon mode (de)multiplexer enabling high capacity photonic networks-on-chip with a single-wavelength carrier light, Opt. Lett. 38 (2013) 1422-1424.

[15] J. Wang, Y. Xuan, M. Qi, H. Huang, Y. Li, M. Li, X. Chen, Z. Sheng, A. Wu, W. Li, X. Wang S. Zou, F. Gan, Broadband and fabrication-tolerant on-chip scalable mode-division multiplexing based on mode-evolution counter-tapered couplers, Opt. Lett. 40 (2015) 1956-1959.

[16] T. Uematsu, Y. Ishizaka, Y. Kawaguchi, K. Saitoh, M. Koshiba, Design of a compact two-mode multi/demultiplexer consisting of multimode interference waveguides and a wavelength-insensitive phase shifter for mode-division multiplexing transmission, J. Lightwave Technol. 30 (2012) 2421-2426.

[17] Y. Li, C. Li, C. Li, B. Cheng, C. Xue, Compact two-mode (de) multiplexer based on symmetric Y-junction and multimode interference waveguides, Opt. Express 22 (2014) 5781-5786.

[18] L. F. Frellsen, Y. Ding, O. Sigmund, L. H. Frandsen, Topology optimized mode multiplexing in silicon-on-insulator photonic wire waveguides, Opt. Express 24 (2016) 16866-16873.

[19] Z. Jafari, A. Zarifkar, M. Miri, Compact fabrication-tolerant subwavelength-grating-based two-mode division (de)multiplexer, Appl. Opt. 56 (2017) 7311-7319.

[20] W. Jiang, J. Hu, S. Mao, H. Zhang, L. Zhou, B. M. A. Rahman, Broadband silicon four-mode (de)multiplexer using subwavelength grating-assisted triple-waveguide couplers, J. Lightwave Technol. 39 (2021) 5042-5047.

[21] Y. He, Y. Zhang, Q. Zhu, S. An, R. Cao, X. Guo, C. Qiu, Y. Su, Silicon high-order mode (de)multiplexer on single polarization, J. Lightwave Technol. 36 (2018) 5746-5753.

[22] D. González-Andrade, J. G. Wangüemert-Pérez, A. V. Velasco, A. Ortega-Moñux, A. Herrero-Bermello, I. Molina-Fernández, R. Halir, P. Cheben, Ultra-broadband mode converter and multiplexer based on sub-wavelength structures, IEEE Photonics J. 10 (2018) 2201010.

[23] D. González-Andrade, A. Dias, J. G. Wangüemert-Pérez, A. Ortega-Moñux, Í. Molina-Fernández, R. Halir, P. Cheben, A. V. Velasco, Experimental demonstration of a broadband mode converter and multiplexer based on subwavelength grating waveguides, Opt. Laser Technol. 129 (2020) 106297.

[24] L. Xu, Y. Wang, D. Mao, J. Zhang, Z. Xing, E. El-Fiky, Md. G. Saber, A. Kumar, Y. D'Mello, M. Jacques, D. V. Plant, Ultra-broadband and compact two-mode multiplexer based on subwavelength-grating-slot-assisted adiabatic coupler for the silicon-on-insulator platform, J. Lightwave Technol. 37 (2019) 5790-5800.

[25] L. Zhang, J. Xiao, Compact and broadband mode demultiplexer using a subwavelength grating engineered MMI coupler, J. Opt. Soc. Am. B 38 (2021) 2830-2836.

[26] D. González-Andrade, R. Fernández de Cabo, J. Vilas, I. Olivares, A. Dias, J. M. Luque-González, J. G. Wangüemert-Pérez, A. Ortega-Moñux, Í. Molina-Fernández, R. Halir, P. Cheben, A. V. Velasco, Mode converter and multiplexer with a subwavelength phase shifter for extended broadband operation, IEEE Photonics Technol. Lett. 33 (2021) 1262-1265.

[27] J. Wang, P. Chen, S. Chen, Y. Shi, D. Dai, Improved 8-channel silicon mode demultiplexer with grating polarizers, Opt. Express 22 (2014) 12799.

[28] A. T. Tran, D. C. Truong, H. T. Nguyen, Y. V. Vu, A new simulation design of three-mode division (de)multiplexer based on a trident coupler and two cascaded 3×3 MMI silicon waveguides, Opt. Quantum Electron. 50 (2018) 426.

[29] C. D. Truong, T. H. Nguyen, Q. T. Pham, M. T. Trinh, K. Vu, Three-mode multiplexer and demultiplexer utilizing trident and multimode couplers, Opt. Commun. 435 (2019) 334-340.

[30] Z. Wang, C. Yao, Y. Zhang, Y. Su, Compact silicon three-mode multiplexer by refractive-index manipulation on a multi-mode interferometer, Opt. Express 29 (2021) 13899-13907.

[31] Z. L. Hussain, R. S. Fyath, Design and simulation of a compact three-mode (de)multiplexer based on a subwavelength grating multimode interference coupler, Photonics Nanostructures – Fudam. Appl. 47 (2021) 100966.

[32] Z. L. Hussain, R. S. Fyath, Design and simulation of 4-mode (de)multiplexers implemented in conventional and subwavelength grating Si/SiO$_2$ platforms, Optik 251 (2022) 168449.

[33] L. Lu, D. Liu, M. Yan, M. Zhang, On-chip mode converter based on cross-connected subwavelength Y-junctions, Photonics Res. 9 (2021) 43-48.

[34] M. Bachmann, P. A. Besse, H Melchior, General self-imaging properties in N×N multimode interference couplers including phase relations, Appl. Opt. 33 (1994) 3905-3911.

[35] P. Cheben, R. Halir, J. H. Schmid, H. A. Atwater, D. R. Smith, Subwavelength integrated photonics, Nature 560 (2018) 565-572.

[36] P. Cheben, D.-X. Xu, S. Janz, A. Densmore, Subwavelength waveguide grating for mode conversion and light coupling in integrated optics, Opt. Express 14 (2006) 4695-4702.

[37] R. Halir, P. J. Bock, P. Cheben, A. Ortega-Moñux, C. Alonso-Ramos, J. H. Schmid, J. Lapointe, D.-X. Xu, J. G. Wangüemert-Pérez, Í. Molina-Fernández, S. Janz, Waveguide sub-

wavelength structures: a review of principles and applications, Laser Photonics Rev. 9 (2015) 25-49.

[38] J. M. Luque-González, A. Sánchez-Postigo, A. Hadij-ElHouati, A. Ortega-Moñux, J. G. Wangüemert-Pérez, J. H. Schmid, P. Cheben, Í. Molina-Fernández, R. Halir, A review of silicon subwavelength gratings: building break-through devices with anisotropic metamaterials, Nanophotonics 10 (2021) 2765-2797.

[39] R. Halir, P. Cheben, J. M. Luque-González, J. D. Sarmiento-Merenguel, J. H. Schmid, G. Wangüemert-Pérez, D.-X. Xu, S. Wang, A. Ortega-Moñux, Í. Molina-Fernández, Ultra-broadband nanophotonic beamsplitter using an anisotropic sub-wavelength metamaterial, Laser Photonics Rev. 10 (2016) 1039-1046.

[40] L. Xu, Y. Wang, D. Patel, M. Morsy-Osman, R. Li, M. Hui, M. Parvizi, N. Ben-Hamida, D. V. Plant, Ultra-broadband and ultra-compact optical 90° hybrid based on 2×4 MMI coupler with subwavelength gratings on silicon-on-insulator, Optical Fiber Communication Conference (2018), p. M3I.7.

[41] D. González-Andrade, J. M. Luque-González, J. G. Wangüemert-Pérez, A. Ortega-Moñux, P. Cheben, Í. Molina-Fernández, A. V. Velasco, Ultra-broadband nanophotonic phase shifter based on subwavelength metamaterial waveguides, Photonics Res. 8 (2020), 359-367.

[42] R. Fernández de Cabo, D. González-Andrade, P. Cheben, A. V. Velasco, High-performance on-chip silicon beamsplitter based on subwavelength metamaterials for enhanced fabrication tolerance, Nanomaterials 11 (2021) 1304.

[43] L. Lu, D. Liu, M. Yan, M. Zhang, Subwavelength adiabatic multimode Y-junctions, Opt. Lett. 44 (2019) 4729-4732.

[44] R. Fernández de Cabo, J. Vilas, P. Cheben, A. V. Velasco, D. González-Andrade, Experimental characterization of an ultra-broadband dual-mode symmetric Y-junction based on metamaterial waveguides, Opt. Laser Technol. 157 (2023) 108742.